\documentclass[aps,prl,reprint,groupedaddress,amsmath,amssymb,longbibliography,floatfix]{revtex4-2}

\usepackage{graphicx}
\usepackage{dcolumn}
\usepackage{bm}
\usepackage{xcolor}
\usepackage{epsf}
\usepackage{epsfig}

\begin{document}

\title{Experimental fingerprint of the electron's longitudinal momentum\\ at the tunnel exit in strong field ionization}

\author{A. Geyer$^1$}
\email{geyer@atom.uni-frankfurt.de}
\author{D. Trabert$^1$}
\author{M. Hofmann$^1$}
\author{N. Anders$^1$}
\author{M. S. Schöffler$^1$}
\author{L. Ph. H. Schmidt$^1$}
\author{T. Jahnke$^2$}
\author{M. Kunitski$^1$}
\author{R. Dörner$^1$}
\author{S. Eckart$^1$}
\email{eckart@atom.uni-frankfurt.de}

\affiliation{$^1$ Institut f\"ur Kernphysik, Goethe-Universit\"at, Max-von-Laue-Str. 1, 60438 Frankfurt am Main, Germany}
\affiliation{$^2$ European XFEL, 22869 Schenefeld, Germany}

\date{\today}
\begin{abstract}
We present experimental data on the strong field tunnel ionization of argon in a counter-rotating two-color (CRTC) laser field. We find that the initial momentum component along the tunneling direction changes sign comparing the rising and the falling edge of the CRTC field. If the initial momentum at the tunnel exit points in the direction of the ion at the instant of tunneling, this manifests as an enhanced Coulomb interaction of the outgoing electron with its parent ion. Our conclusions are in accordance with predictions based on strong field approximation.
\end{abstract}

\maketitle

Tunnel ionization \cite{Keldysh1965} can occur when a single atom or molecule is exposed to a strong laser field. This process is often modeled as a two-step process, in which a bound electron tunnels through the barrier that is formed by the superposition of the laser and the ionic potential. Subsequently, the electron is driven by the time-dependent laser field in the presence of the ionic potential \cite{Corkum1993}. Many processes, such as high harmonic generation \cite{krauseetal_cutofflaw,Becker1999,fleischer2014spin}, non-sequential double ionization \cite{Walker1994_Helium,Eckart2016,Mancuso2016PRL} and laser-induced electron diffraction \cite{zuo1996laser, Meckel2008, blaga2012imaging} rely on an accurate modeling of tunneling dynamics.

For a static electric field, tunneling occurs adiabatically. In this special case, the momentum distribution upon tunneling is typically assumed to be perpendicular to the tunneling direction and described by a two-dimensional Gaussian distribution that is centered at zero momentum \cite{Keldysh1965,Xu2018}. However, the final momentum distributions can be cusp-like due to Coulomb focusing \cite{Rudenko2005,Brabec1996,Eckart2018SubCycle}. In contrast to adiabatic tunneling, the description of tunneling in a time-dependent laser electric field is more intricate, since the tunnel exit position moves in position space and the speed of this movement affects the dynamics in the tunnel. Tunneling under these conditions is referred to as non-adiabatic tunneling \cite{Perelomov1966,Yudin2001A,Misha2005,Boge2013,Eckart2018_Offsets}. For non-adiabatic tunneling there can be a momentum component pointing along the tunneling direction, which is called longitudinal momentum. The width of the longitudinal momentum distribution has been investigated theoretically \cite{Tian2017, Xu2018} and experimentally \cite{Pfeiffer_2012, hofmann2013comparison, Xufei2014, Li2019}. Recently, theoretical studies have predicted a sign change of the longitudinal momentum distribution as a function of the sub-cycle laser phase \cite{Han2017b,Luo2019,Xiao_2022}.

In this Letter, we experimentally observe two qualitative features in the electron momentum distribution that emerge because of the longitudinal momentum and which become prominent for an appropriately chosen two-color laser field. The key idea of our approach is, that an initial momentum along or opposite to the tunnel direction alters the electron's interaction with the Coulomb potential of the left behind ion. This Coulomb effect manifests itself in two different ways: Firstly it gives rise to a narrowing of the momentum distribution along the light-pro-
\onecolumngrid

\begin{figure}[ht]
\includegraphics[width=\textwidth]{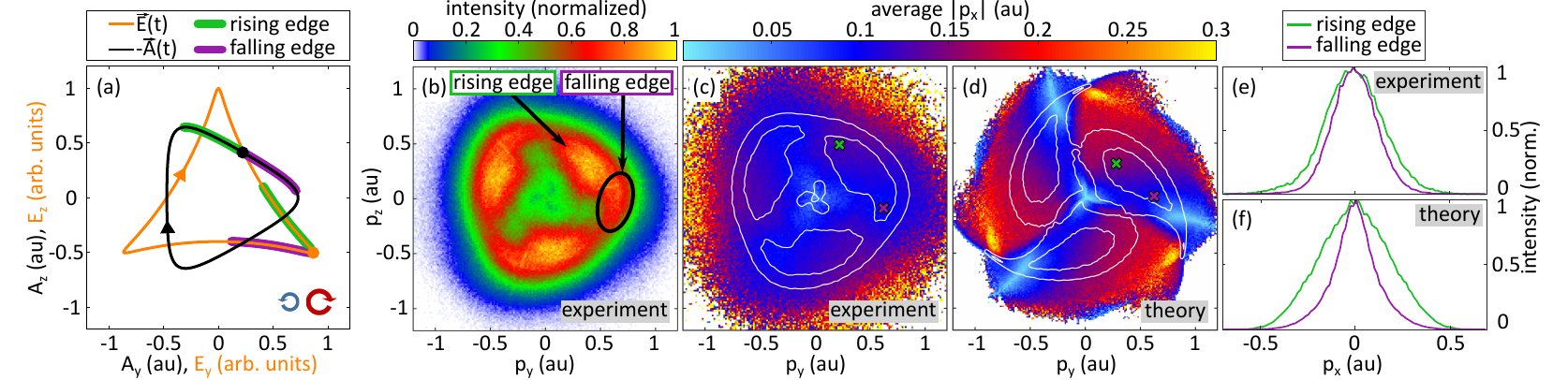}
\caption{\label{fig1} Electron momentum distribution for the strong field ionization in a counter-rotating two-color (CRTC) field. (a) electric laser field $\vec{E}(t)$ (orange) and the corresponding negative vector potential $-\vec{A}(t)$ (black). Arrows indicate the temporal evolution. The orange dot marks one peak of the laser electric field $|\vec{E}(t_0)|$ and the black dot marks $-\vec{A}(t_0)$. The rising [falling] edge of the CRTC field and the corresponding vector potential are indicated. (b) Measured electron momentum distribution for the CRTC field shown in (a). The black circle marks the wing-like structure of one of the peaks. The average of $|p_x|$, which is a measure for the width of the momentum distribution in light-propagation direction, is shown as obtained by experiment (c) and theory (d). The white lines in (c) [(d)] are contour lines for the distribution shown in (b) [2(a)]. (e) [(f)] shows the $p_x$-distributions for the points in the $p_yp_z$-plane that are indicated by crosses in (c) [(d)].}
\end{figure}
\twocolumngrid

\noindent pagation direction, which can even lead to a cusp-like feature \cite{Rudenko2005,Brabec1996,Eckart2018SubCycle}. Secondly, the Coulomb potential leads to a rotation of the final momentum distribution in the polarization plane, which is closely related to the attoclock angle \cite{Bray2018,Eckle2008,sainadh2019attosecond,trabert2021angular}. 

In Fig. \ref{fig1} we present data on the single ionization of argon by a counter-rotating two-color (CRTC) field. Fig. \ref{fig1}(a) shows the laser electric field and the corresponding negative vector potential. Fig. \ref{fig1}(b) shows the measured electron momentum distribution in the polarization plane ($p_yp_z$ plane) integrated over the momentum in light-propagation direction $p_x$. Electrons that are born on the falling edge of the CRTC field form a wing-like structure (see Fig. \ref{fig1}(b)). Here, we mean by ``falling edge'' the time interval on a sub-cycle time scale, in which the magnitude of the laser electric field decreases as a function of time (see Fig. \ref{fig1}(a)). The wing-like structure is the first qualitative feature of an enhanced Coulomb effect. The second observable that reveals an enhanced Coulomb effect for electrons that are born on the falling edge of the CRTC field is the width of the $p_x$-distribution. Fig. \ref{fig1}(c) shows the average of $|p_x|$, which is a measure of the width of the $p_x$-distribution. The wing-like structure possesses small values for the average of $|p_x|$, which underlines the existence of an enhanced Coulomb effect for these electrons (purple cross in Fig. \ref{fig1}(c)). In previous studies, such narrow distributions along $p_x$ were observed for low energy electrons only \cite{Rudenko2005, Brabec1996, Eckart2018SubCycle} and were due to Coulomb focusing. 

Thus, the measured three-dimensional electron momentum distribution shows two manifestations of an enhanced Coulomb effect for electrons that are born at the falling edge of the CRTC field as compared to electrons that are born on the rising edge. We will show that this difference is caused by a sub-cycle modulation of the initial momenta in the tunneling direction. In the remainder of this Letter we will provide experimental details and support our interpretation with trajectory based simulations using a state-of-the-art method to model non-adiabatic tunnel ionization. The essence of those simulations is shown in Fig. \ref{fig1}(d) and \ref{fig1}(f). In particular Fig. \ref{fig1}(f) shows a narrower $p_x$-distribution for electrons born on the falling edge of the laser electric field than electrons born on the rising edge. This agrees with our experimental observation (see Fig. \ref{fig1}(e)).

\begin{figure}[t]
\includegraphics[width=\columnwidth]{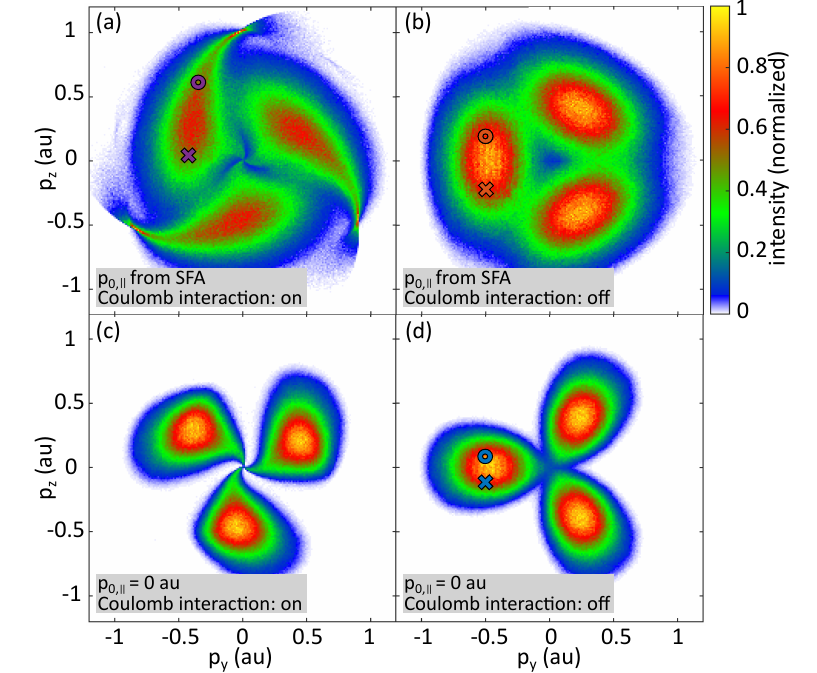}
\caption{\label{fig2} Simulated electron momentum distributions from the non-adiabatic classical two-step (NACTS) model. (a) shows the simulated electron momentum distribution for the laser field shown in Fig. \ref{fig1}(a). (b) shows the same as (a) but here the Coulomb interaction after tunneling is neglected. (c) and (d) show the same as (a) and (b) but the initial momentum along the direction of the laser electric field at the instant of tunneling $p_{0,\parallel}$ is set to zero ($p_{0,\parallel}=0$\,au). The crosses [circles] in (a), (b) and (d) mark the final momenta of the trajectories that are shown in Fig. \ref{fig4}(a) [(b)].}
\end{figure}

The CRTC field is created as the coherent superposition of two laser pulses with central wavelengths of 780 nm and 390 nm. Both pulses are circularly polarized and have opposite helicities. The intensity of the 780\,nm [390\,nm] pulse is $8.5 \times 10^{13}$\,W/cm$^2$ [$5.6 \times 10^{13}$\,W/cm$^2$]. We used an interferometric two-color laser setup to generate this CRTC field. To this end, 40-fs laser pulses with a repetition rate of 8\,kHz (laser system ``Dragon'' by KMLabs) and a central wavelength of 780\,nm are frequency doubled in a $\beta$-barium borate crystal to generate the second harmonic (390\,nm). The two laser pulses at different wavelengths are spatially separated by a beamsplitter. This allows us to tailor the polarization and the intensity of both pulses individually. The relative phase of the two single-color laser pulses can be adjusted by a nanometer-delay stage before they are merged using a beam combiner. The optical setup is the same as in \cite{Eckart2018_Offsets,EckartNatPhys2018,Eckart2016}. Inside a cold target recoil ion momentum spectroscopy (COLTRIMS) reaction microscope \cite{trabert2021angular} we use a spherical mirror ($f=80$\,mm) to focus the CRTC field onto a jet of argon that is created by supersonic gas expansion into vacuum. After ionization the charged fragments are guided by a homogeneous electric field of 18\,V/cm and a magnetic field of 10\,G to position- and time-sensitive detectors \cite{jagutzki2002multiple}. This allows us to measure the three-dimensional electron and ion momenta in coincidence \cite{ullrich2003recoil}. During the measurement the relative phase of the two colors was actively varied. These variations and additional slow drifts of the relative phase in the interferometer are compensated in the offline data analysis as in Ref. \cite{Eckart2018_Offsets}. The intensity calibration was done \textit{in situ} for both colors separately and takes volume averaging into account. For the single color field at 780\,nm the electron drift momentum from the ionization of argon by circularly polarized light is used to calibrate the intensity \cite{Eckart2018_Offsets}. For the intensity calibration of the single color field at 390\,nm the shift of the above threshold ionization peaks from the ionization of argon by circularly polarized light is analyzed as a function of the light's intensity (as in Ref. \cite{Eckart2016}).

We have performed a simulation using the non-adiabatic classical two-step (NACTS) model to reproduce the observables that reveal an enhanced Coulomb effect. The NACTS model \footnote{The NACTS model is the same as in \cite{Trabert2021Atomic}. Accordingly, the NACTS is also equivalent to the model presented in \cite{Brennecke2020_gouy}, except that the phase of trajectories after tunneling is neglected in the NACTS model.} is a combination of strong field approximation (SFA) \cite{Popruzhenko2008,Yan_2012} and the classical two-step (CTS) model \cite{Shilovski2016}. This combination is very versatile since the NACTS model includes non-adiabatic dynamics in the classically forbidden region (tunnel) like non-adiabatic momentum offsets perpendicular and parallel to the tunneling direction \cite{Olga2011A,Han2017b,Eckart2018_Offsets}. The CTS model accounts for the propagation of the electron after tunneling in the combined potential of the ion and the laser electric field. Technically, in the NACTS simulation we use the SFA to model the dynamics in the classically forbidden region. After the electron leaves this region its properties (instant of tunneling, tunnel exit position, initial momentum and tunneling probability) are used as initial conditions for the CTS model. The time-dependent laser field, Coulomb interaction and initial conditions from SFA determine the classical electron trajectories that eventually lead to the final momenta shown in Fig. \ref{fig2}(a). It is remarkable that the NACTS model reproduces both fingerprints of the longitudinal momentum: the wing-like structure and a narrowed distribution along $p_x$ for the falling edge of the laser field (see Fig. \ref{fig1}(b)-(f) and Fig. \ref{fig2}(a)).

Fig. \ref{fig2} shows the final electron momentum distributions from the NACTS model for four different cases to investigate the interplay of Coulomb interaction and the longitudinal momentum at the tunnel exit. In Fig. \ref{fig2}(a) the full NACTS simulation has been used. Next, we take the NACTS model but switch off Coulomb interaction after tunneling. This leads to the result shown in Fig. \ref{fig2}(b). Here, the two sides of the maximum,  that belong to the rising and falling edge of the CRTC field are symmetrical. In particular, the electron momentum distribution would not change upon inversion of the direction of rotation of the CRTC field in this case. For Fig. \ref{fig2}(c), we include Coulomb interaction after tunneling but set the electron momentum parallel to the laser electric field at the instant of tunneling $p_{0,\parallel}$ to zero. Effectively, this ensures that the electron's initial momentum is solely perpendicular to the laser electric field when the electron leaves the tunnel (as it is for adiabatic tunneling). Compared to Fig. \ref{fig2}(b) we see that it is the Coulomb interaction after tunneling that induces the sense of rotation to the final momentum distribution in Fig. \ref{fig2}(c). In particular, we observe that slow electrons form a spiral pattern. In comparison, fast electrons experience an angular offset \cite{Bray2018,Eckle2008,sainadh2019attosecond,trabert2021angular}, but there is no wing-like structure as in Fig. \ref{fig2}(a) and Fig. \ref{fig1}(b).  In a final variation of our model, we switch off Coulomb interaction after tunneling and also set $p_{0,\parallel}$ to zero. This gives rise to the result in Fig. \ref{fig2}(d). Interestingly, neither Fig. \ref{fig2}(b) nor Fig. \ref{fig2}(c) shows a wing-like structure or a sense of rotation for high electron energies. This indicates that both $p_{0,\parallel}$ and Coulomb interaction after tunneling are essential to reproduce the wing-like structure (compare Fig. \ref{fig1}(b) and Fig. \ref{fig2}(a)) and the narrowed distribution along $p_x$ for the falling edge of the laser field (see Fig. \ref{fig1}(e) and \ref{fig1}(f)).

\begin{figure}[t!]
\includegraphics[width=\columnwidth]{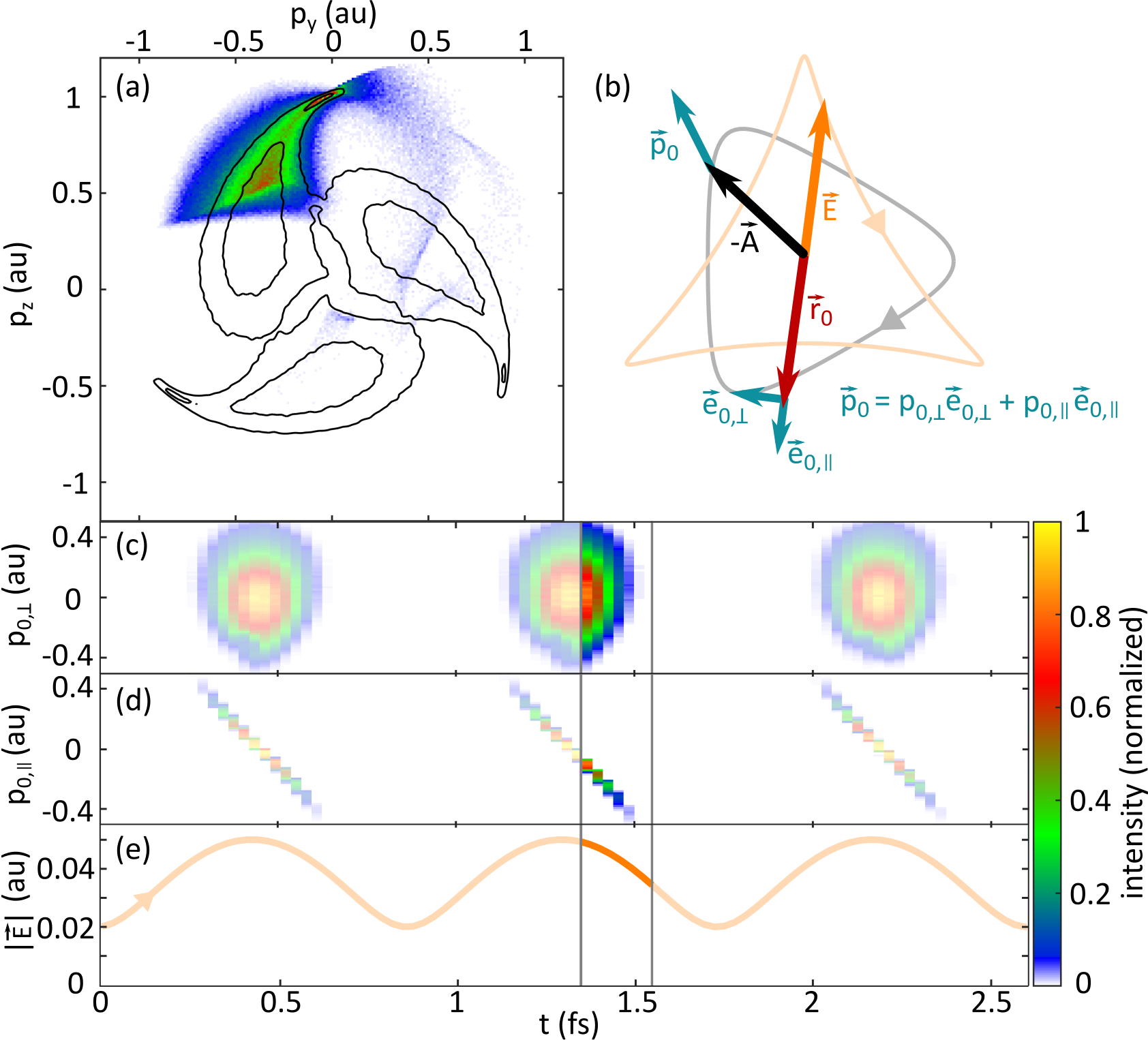}
\caption{\label{fig3} Theoretical modeling of the ionization process. The black lines in (a) show contour lines for the distribution shown in Fig. \ref{fig2}(a). (b) illustrates the geometry of the initial momenta for an instant of tunneling $t=$ 1.5\,fs: The laser electric field $\vec{E}$, the negative vector potential $-\vec{A}$, the tunnel exit $\vec{r}_0$ and the initial momentum $\vec{p}_0$ are depicted. The turquoise arrows define the unit vectors for the electron's initial momentum parallel and perpendicular to the laser electric field $\vec{e}_{0,\parallel}$ and $\vec{e}_{0,\perp}$. (c) [(d)] shows the initial momentum perpendicular [parallel] to the time-dependent laser electric field at the instant of tunneling $p_{0,\perp}$ [$p_{0,\parallel}$] as a function of $t$. (e) shows $|\vec{E}|$ as a function of $t$. The initial momenta and the laser electric field between the two gray lines in (c)-(e) belong to final momenta that are shown as the colored distribution in (a).}
\end{figure}

Fig. \ref{fig3} shows details on the sub-cycle dependence of the longitudinal momentum at the tunnel exit within the NACTS model. The colored data in Fig. \ref{fig3}(a) shows the subset of modeled trajectories that are released from the tunnel during the falling edge of the CRTC field. This corresponds to electron release times of 1.35\,fs $<t<$ 1.55\,fs (see gray lines in Fig. \ref{fig3}(c)-(e)). The coordinate system and the tunneling geometry is illustrated in Fig. \ref{fig3}(b) for $t=$ 1.5\,fs. It should be noted, that the laser electric field  and the negative vector potential are not perpendicular at this time. Fig. \ref{fig3}(c) shows the initial momentum perpendicular to the laser field at the instant of tunneling $p_{0,\perp}$ as a function of $t$. For each value of $t$, $p_{0,\perp}$ shows the well-known Gaussian-like distribution as obtained via SFA. The $p_{0,\perp}$-distribution is slightly shifted towards positive values due to the non-adiabatic initial momentum offset \cite{Eckart2018_Offsets}. Further, $p_{0,\perp}$ is the same comparing the rising and the falling edge of the laser pulse (see $|\vec{E}|$ in Fig. \ref{fig3}(e)). Fig. \ref{fig3}(d) depicts $p_{0,\parallel}$ as obtained via SFA. There are three sharp lines with negative slope. This indicates that $p_{0,\parallel}$ has positive values on the rising edge of the electric field, is zero for the peak electric field and has negative values on the falling edge. It is important to note that for negative $p_{0,\parallel}$ the electron's momentum vector at the tunnel exit has a component pointing back to the ionic core. 

\begin{figure}[t]
\includegraphics[width=\columnwidth]{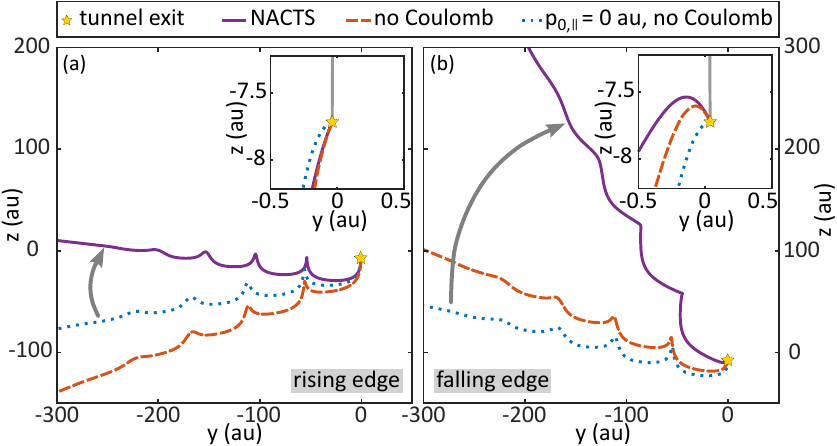}
\caption{\label{fig4} Characteristic electron trajectories from the NACTS simulation. (a) [(b)] shows trajectories that are released on the rising [falling] edge of the laser electric field. The inset shows a zoom to the position coordinates at the tunnel exit. The purple lines are typical trajectories within the NACTS model. The red dashed lines show trajectories with $p_{0,\parallel}$ from SFA that neglect Coulomb interaction after tunneling. The blue dotted lines show trajectories without Coulomb interaction and with $p_{0,\parallel}=0$\,au. The star marks the tunnel exit position and the gray line in the insets connects the tunnel exit position with the ion's location (origin of the coordinate system). The grey arrows indicate the offset angle caused by the interplay of $p_{0,\parallel}$ and Coulomb interaction after tunneling. It is evident that this offset angle is larger for the falling edge of the laser electric field than it is for the rising edge. The final electron momenta of the trajectories that are shown in (a) [(b)] are indicated with crosses [circles] in Fig. \ref{fig2}.}
\end{figure}

Fig. \ref{fig4} depicts characteristic electron trajectories from the NACTS simulation in position space. Fig. \ref{fig4}(a) shows three electron trajectories that are released on the rising edge of the laser electric field. The star marks the tunnel exit position as obtained from SFA. Here, the tunnel length from SFA is slightly shorter than in the TIPIS (tunnel
ionization in parabolic coordinates with induced dipole and Stark shift) model \cite{Pfeiffer2012naturalcoordinates}. It should be noted, that the tunnel length decreases as a function of $p_{0,\perp}$ as shown in \cite{Trabert2021Atomic}. This leads to an increased Coulomb effect for high final radial electron momenta, which is also in line with the findings from \cite{Trabert2021Atomic} and additionally enhances the Coulomb effect for the experimentally observed wing-like structure (see Fig. \ref{fig1}(b)). The purple line in Fig. \ref{fig4}(a) shows a typical trajectory from the NACTS model for the rising edge of the CRTC field. The corresponding final electron momentum is indicated by a purple cross in Fig. \ref{fig2}(a). If one neglects Coulomb interaction after tunneling, this modifies the trajectory (shown as red dashed line in Fig. \ref{fig4}(a)) and results in a slightly different final momentum (see red cross in Fig. \ref{fig2}(b)). If we additionally set $p_{0,\parallel}$ to zero this leads to the blue dotted trajectory in Fig. \ref{fig4}(a) (see blue cross in Fig. \ref{fig2}(d)). In full analogy, Fig. \ref{fig4}(b) shows trajectories that are released on the falling edge of the laser electric field. The grey arrows in Fig. \ref{fig4}(a) and \ref{fig4}(b) illustrate the change of emission direction of the position space trajectory, that is due to the interplay of $p_{0,\parallel}$ and  Coulomb interaction after tunneling. It is evident that this emission angle is much more affected by Coulomb interaction on the falling edge than it is affected on the rising edge. This results in an asymmetric smearing out of the angular maxima in one direction (clockwise direction in our case) and thereby gives rise to the wing-like structure (see Fig. \ref{fig1}(b) and Fig. \ref{fig2}(a)).

The reported sub-cylce sign change of the longitudinal momentum is in accordance with previous predictions based on SFA \cite{Han2017b,Luo2019,Xiao_2022}. The present work shows that this theoretical prediction regarding the longitudinal momentum is also experimentally accessible since it manifests itself as an enhanced Coulomb effect. It might be counterintuitive that electrons escape out of the tunnel even though they have an inward directed momentum. This pseudo-problem is eased by the fact that the so-called tunnel exit is the starting point of the classical trajectory. Therefore inward directed initial momenta lead to trajectories which start uphill the potential barrier, are slowed down to zero and are then finally driven outwards.

In conclusion, we have studied the strong field ionization of argon in a tailored CRTC field. The measured electron momentum distribution exhibits a wing-like structure that is due to the Coulomb effect and results in a substantial narrowing of the momentum distribution in the light-propagation direction. Our experimental data shows that the initial momentum along the tunnel direction points away from [towards] the ionic core for the rising [falling] edge of the laser electric field as predicted by SFA \cite{Han2017b, Xiao_2022}. Consequently, Coulomb interaction is strongly enhanced for electrons that are released on the falling edge of the laser field. We expect that the agreement of experiment and theory could be further improved in future theoretical studies by taking the Coulomb potential in the classically forbidden region (tunnel) into account. To this end, our results serve as a benchmark for future theoretical studies that aim at the inclusion of Coulomb interaction in the tunnel \cite{Eckle2008,sainadh2019attosecond,Trabert2021Atomic}. Finally, our approach establishes experimental access to the sign of the longitudinal momentum of the electron at the tunnel exit.

\section{Acknowledgments}
\normalsize
The experimental work was supported by the DFG (German Research Foundation). S.E. acknowledges funding of the DFG through Priority Programme SPP 1840 QUTIF. We acknowledge support from Deutsche Forschungsgemeinschaft via Sonderforschungsbereich 1319 (ELCH).

\end{document}